\documentclass[conference]{IEEEtran}

\usepackage[linesnumbered,ruled,vlined]{algorithm2e}

\IEEEoverridecommandlockouts
\usepackage{amsmath,amssymb,amsfonts}
\usepackage{algorithmic}
\usepackage{graphicx}
\usepackage{textcomp}
\usepackage{xcolor}
\usepackage{subfig}
\usepackage{diagbox}

\usepackage[linesnumbered,ruled,vlined]{algorithm2e}
\usepackage{booktabs}
\usepackage{multirow}
\def\BibTeX{{\rm B\kern-.05em{\sc i\kern-.025em b}\kern-.08em
    T\kern-.1667em\lower.7ex\hbox{E}\kern-.125emX}}
\begin{document}

\title{ Single Channel-based Motor Imagery Classification using Fisher's Ratio and Pearson Correlation

\thanks{This publication has emanated from research supported in part by a grant from Science Foundation Ireland under Grant numbers 18/CRT/6183, 18/CRT/6224, SFI/12/RC/2289\_P2. For the purpose of Open Access, the author has applied a CC BY public copyright licence to any Author Accepted Manuscript version arising from this submission.}
}

\author{\IEEEauthorblockN{Sonal Santosh Baberwal}
\IEEEauthorblockA{\textit{School of Electronic Engineering} \\
\textit{Dublin City University}\\
Dublin, Ireland \\
sonal.baberwal2@mail.dcu.ie}
\and

\IEEEauthorblockN{Tomas Ward}
\IEEEauthorblockA{\textit{School of Computing} \\
\textit{Dublin City University}\\
Dublin, Ireland \\
tomas.ward@dcu.ie}
\and
\IEEEauthorblockN{Shirley Coyle}
\IEEEauthorblockA{\textit{School of Electronic Engineering} \\
\textit{Dublin City University}\\
Dublin, Ireland \\
shirley.coyle@dcu.ie}

}

\maketitle

\begin{abstract}
Motor imagery-based BCI systems have been promising and gaining popularity in rehabilitation and Activities of daily life(ADL). Despite this, the technology is still emerging and has not yet been outside the laboratory constraints. Channel reduction is one contributing avenue to make these systems part of ADL. Although Motor Imagery classification heavily depends on spatial factors, single channel-based classification remains an avenue to be explored thoroughly. Since Fisher's ratio and Pearson Correlation are powerful measures actively used in the domain, we propose an integrated framework (FRPC integrated framework) that integrates Fisher's Ratio to select the best channel and Pearson correlation to select optimal filter banks and extract spectral and temporal features respectively. The framework is tested for a 2-class motor imagery classification on 2 open-source datasets and 1 collected dataset and compared with state-of-art work. Apart from implementing the framework, this study also explores the most optimal channel among all the subjects and later explores classes where the single-channel framework is efficient.

\end{abstract}


\section{Introduction}

A Brain-Computer Interface (BCI) system establishes a direct communication pathway between the brain and an external device or computer system \cite{nicolas2012brain},\cite{wolpaw2002brain}. BCIs enable individuals to control devices, interact with software, or communicate without relying on traditional neuromuscular pathways such as peripheral nerves and muscles. EEG is widely used in BCIs due to its non-invasiveness and high temporal resolution. The recorded EEG signals are processed to extract features related to brain activity, which are then interpreted by a computer algorithm to derive user commands. These systems are mainly classified depending on the type of the system i.e. Event-Related Potentials (ERPs) and Motor Imagery.
   
Motor imagery, i.e. imagination of the movements of motor parts \cite{pfurtscheller1997motor}, is a cognitive process involving the mental simulation of movement, and has gained prominence in various applications, particularly in brain-computer interface (BCI) systems. The ability to decode motor intentions from electroencephalography (EEG) signals holds immense potential for developing personalized and efficient neurofeedback systems/ BCI systems. The BCI system processes these EEG signals, classifying them into specific motor commands\cite{lotte2014tutorial}. The personalized nature of motor imagery allows for subject-specific customization, enhancing the accuracy and efficiency of the BCI system. Common applications include neurorehabilitation, prosthetic control, and assistive technology. They enable individuals recovering from neurological disorders or injuries to engage in mental exercises that simulate physical movements, promoting neural plasticity and aiding in the restoration of motor functions. Recent trends in applications show significant work for Stroke Rehabilitation, especially with upper limb/hand exoskeletons. Significant improvements have been reported when the patients have been using the hand exoskeletons controlled by their MI signals \cite{frolov2017post}, thereby extending MI-based BCI systems not only with medical applications such as Spinal-cord injuries (SCI) but also with Stroke Rehabilitation. Thus, MI-based BCI systems show potential for Neuroprosthetic BCI systems and rehabilitation to move out of research laboratories and extend the application of motor imagery-based BCIs to assist individuals in performing ADL. Integrating Brain-Computer Interface (BCI) systems into daily routines \cite{stolzle2024guiding}, particularly in uncontrolled environments, is indeed a complex task. Among the challenges faced, issues related to device portability, number of channels, user acceptance, and the need for continuous improvement in system robustness stand out. Channel reduction emerges as a viable solution, especially when considering the application of BCIs in ADL.

There has been research on single channel-based motor imagery classification, for instance, Ko et al.,2017,  developed a hybrid MI-SSVEP system using a single channel(C3 or C4) for Motor Imagery and SSVEP respectively, using Short Time Fourier Transform(STFT) \cite{ko2017development}. Camacho et al.,2018 developed a real-time single-channel system using STFT and Grey Level Cooccurrence Matrix(GLCM) \cite{camacho2016real}. Kanoga et al.,2018 performed a thorough comparison study on features (majorly GLCM, Log of Variance, Single channel Common Spatial Patterns) and classifiers(LDA,k-NN, GMM, RF, MLP, SVM)  for single channel-based Motor Imagery classification \cite{kanoga2018comparative}.  

Fisher's ratio and Pearson Correlation coefficient (PCC) have been powerful methodologies used in this domain either for channel/band or feature selection. For instance, Thomas et al.,2009 use the Fisher's ratio to select the optimal band from C3 or C4 and pass features from the specific band from all channels for feature extraction to CSP and further classification \cite{thomas2009new}. Thomas et al.,2019 implement the Fisher's ratio to find discriminative channel sets and further extract features from CSP for classification \cite{thomas2019utilizing}. Thomas et al.,2021 implemented subject-specific Fisher's ratio for optimal channel set and optimal band \cite{thomas2021separability}.

Likewise, Dhiman et al., 2023 and Yu et al.,2021 implemented PCC to select optimal channels \cite{dhiman2023electroencephalogram}, \cite{yu2021cross}. Gaur et al.,2021 implemented PCC to select an optimal channel among C3,Cz, and C4, extract information and use it as a reference \cite{gaur2021automatic}. Jusas et al.,2019 combined squared PCC with other techniques using them for feature extraction \cite{jusas2019classification}.  Meng et al.,2023 proposed a framework that mainly includes optimal channel selection based on Pearson correlation coefficient, subband selection based on power spectrum density and subbands and broadband feature selection based on relevance and Fisher's ratio \cite{meng2023optimal}. 

Although the Fisher's ratio has been implemented for channel selection \cite{thomas2019utilizing}, \cite{yang2016subject}; the number of channels varies from subject to subject, and implementation using a single channel yet remains a domain of exploration. Also, a combination of the Fisher's ratio and PCC could possibly lead to efficient features. 

This work delves deeper into the concept of single channel-based classification, thereby proposing an integrated framework using Fisher's Ratio and Pearson Correlation. The framework has been tested on 2 open-source datasets and data collected from 21 participants. Later, we provide a comparison with state-of-art algorithms. Later; the hypothesis of whether the framework performs better on certain classification sets is been explored and the most optimal channel across 39 subjects(combining all datasets used) has been reported. The paper is been concluded with a discussion about the analysis and the results.

\section{Methodology}\label{sec:method}

\subsection{FRPC Integrated Framework}
In the preprocessing phase, initially, a notch filter is applied to eliminate unwanted frequency components from the raw data, such as power line interference, at 50 and 100 Hz. Subsequently, an Independent Component Analysis (ICA) is employed to mitigate artefacts and enhance the separability of motor imagery conditions. The resulting preprocessed data is then focused on three chosen EEG channels, specifically C3, Cz, and C4, which is most popularly known for Motor Imagery \cite{hu2014causality}. 

In our proposed method, we initiate an integrated framework by conducting channel selection using Fisher’s Ratio (FR) and selecting one optimal channel. The primary goal here is to identify the most discriminative EEG channel for subsequent motor imagery classification. Fisher’s Ratio (FR), denoted as \( F R \), serves as our quantitative measure to assess the separability of the data within each channel under all motor imagery conditions.

Mathematically, for a given EEG channel \( i \) without predefined class labels, the Fisher’s Ratio (\( F R_{i} \)) is computed as follows:

\[ F R_{i} = \frac{{(\mu_{i} - \mu_{all})^2}}{{\sigma_{i}^2 + \sigma_{all}^2}} \tag{1}\] 

Here, \( \mu_{i} \) and \( \sigma_{i} \) represent the mean and standard deviation of EEG channel \( i \), while \( \mu_{all} \) and \( \sigma_{all} \) denote the overall mean and standard deviation across all channels, respectively. This ratio emphasizes the between-class variance relative to the within-class variance, aiding in the identification of the optimal channel.

Following the selection of the optimal EEG channel, we advance to sub-band decomposition to capture frequency-specific information. This crucial step involves applying a wavelet transform, a widely used technique for this purpose, to segment the EEG signal into distinct frequency bands. Mathematically, the resulting sub-bands are denoted as \(X_{i,j}\), where \(i\) represents the chosen EEG channel, and \(j\) indexes the sub-band.

The wavelet transform can be expressed as:

\[X_{i,j}(t) = \int_{-\infty}^{\infty} x_i(\tau) \cdot \psi^* \left(\frac{t - \tau}{s_j}\right) d\tau\tag{2}\] 

Here, \(x_i(\tau)\) is the EEG signal from channel \(i\) at time \(\tau\), \(\psi\) is the mother wavelet function, and \(s_j\) denotes the scale parameter related to the frequency band of the sub-band \(j\). The integral captures the convolution of the EEG signal with the scaled and translated wavelet function. The sub-bands are divided into frequency ranges from 8 Hz to 40 Hz with intervals of 4 Hz.

This process results in a set of sub-bands \(X_{i,j}(t)\), each representing a different frequency range. These sub-bands are essential for extracting spectral features relevant to motor imagery conditions, providing a detailed representation of the EEG signal's frequency content.

Within each sub-band, we employ Pearson’s Ratio (PR) to rank the features based on their correlation. Pearson’s Ratio (\( P R_{i,j} \)) for a feature \( x_{i,j} \) in sub-band \( j \) is defined as:

\[ P R_{i,j} = \frac{{(\mu_{x_{i,j}} - \mu_{x_{all,j}})^2}}{{\sigma_{x_{i,j}}^2 + \sigma_{x_{all,j}}^2}}\tag{3} \] 

Here, \( \mu_{x_{i,j}} \) and \( \sigma_{x_{i,j}} \) represent the mean and standard deviation of feature \( x_{i,j} \) within sub-band \( j \), while \( \mu_{x_{all,j}} \) and \( \sigma_{x_{all,j}} \) represent the overall mean and standard deviation of the feature across all channels.

In the final step, the bands are grouped and ranked (n=1(top 1 band),n=2(top 2 bands)...n=6(top 6 bands))in ascending order, and the top 6 bands are selected with high correlation. Later temporal and spectral features from each sub-band are extracted w.r.to their ranking and are fed into an AdaBoost classifier with a crossfold of 10x10; reporting the mean accuracies. This comprehensive framework, integrating Fisher’s Ratio for channel selection, sub-band decomposition, Pearson’s Correlation for feature ranking, and a machine learning classifier, constitutes a robust methodology for subject-specific motor imagery classification. The pipeline, from raw EEG data to classifier input, leverages both spatial and spectral information to enhance classification performance. We employ the proposed framework for a time window of 0.5 to 2.5s; after the event occurs, for every participant and across all datasets. 

\begin{figure}[htbp]
\centering
\includegraphics[width=\columnwidth]{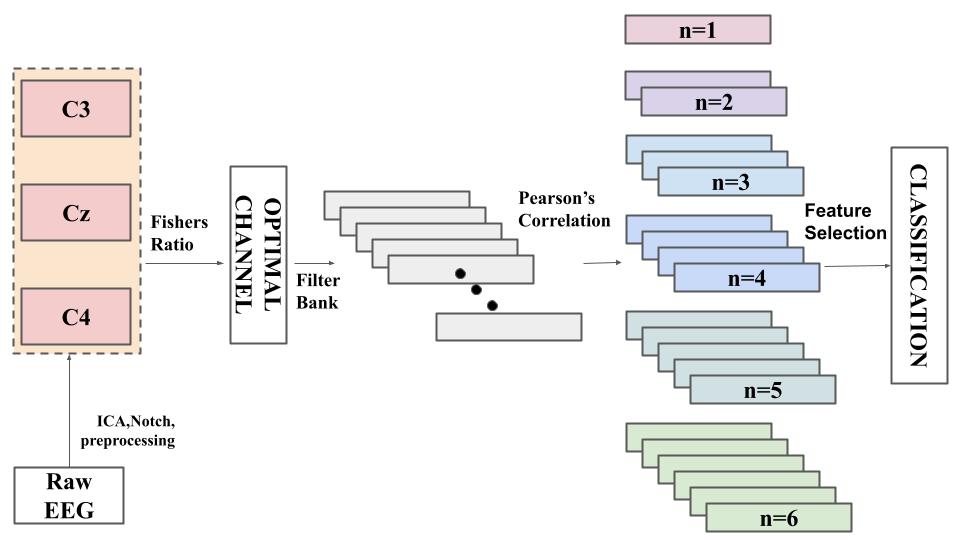} 
\caption{Proposed framework}
\label{fig:protocol}
\end{figure}

\subsection{Dataset}
One of the well-known systems for using  MI to perform respective mental tasks is the Graz paradigm, which reports clinical applications of MI-based BCI systems with operations like controlling a virtual cursor and hand orthosis operations. The Graz paradigm differentiates two or more MI-based brain imagination or  EEG signals synchronously (in predefined time windows) \cite{pfurtscheller1992event}.
To evaluate the proposed methodology, the proposed framework has been tested on the following datasets:

\begin{enumerate}
    \item Dataset 1: The BCI IV-2a dataset encompasses EEG recordings from 9 subjects, employing a cue-based Brain-Computer Interface (BCI) paradigm featuring four distinct motor imagery tasks. Subjects engaged in the mental simulation of movement for the left hand (class 1), right hand (class 2), both feet (class 3), and tongue (class 4). Recorded over two sessions on different days for each participant, each session comprised six runs with short breaks in between. Within each run, 48 trials were conducted, with 12 trials for each of the four possible classes, resulting in a cumulative total of 288 trials per session \cite{brunner2008bci}.
    \item Dataset 2: The BCI IV-2b dataset comprises EEG recordings from 9 healthy participants. Participants performed motor imagery tasks involving left and right-hand movements. Each trial consists of a 6-second baseline, followed by 4 seconds of motor imagery, and annotations indicate the onset of these tasks. The dataset includes three recording training sessions for each participant, providing a total of 316 motor imagery trials across all subjects \cite{leeb2008bci}.
    \item Dataset 3: The data is collected at Dublin City University, as a part of an experiment \cite{baberwal2023protocol}, approved by the Ethical Committee at Dublin City, University, Ireland. The data consists of a total of 21 healthy individuals. The data was collected using the Neuroconcise 3-channel EEG system. The EEG system used is a bipolar system containing seven electrodes, FC3-CP3, FCz-CPz, FC4-CP4 and the bias AFz as ground, with a sampling rate of 125 Hz. The motor imagery training data is recorded in OpenVibe and using Graz-BCI as a reference paradigm \cite{pfurtscheller2003graz} for left and right-hand motor imagery. 
\end{enumerate}
The framework is tested on these datasets and results have been reported. We compare the results with \cite{oikonomou2022multitask} for Dataset 1 and 2, for performance to achieve nearby accuracy and \cite{kanoga2018comparative} for single channel comparison for Dataset 2. We compare Dataset 3 with all channel ERD/S - AdaBoost. Further, all combinations of 2 classes are explored from Dataset 2.  

\section{Results}
For Dataset 1, as seen in Table \ref{tab:2b}, the eLDA framework achieves an impressive average accuracy of 79.12\%. In comparison, the proposed system demonstrates an average accuracy of 62.29\%. The accuracy was high only for 1 participant B03 where the proposed framework achieved an accuracy of  58.12 whereas \cite{oikonomou2022multitask} reports an accuracy of 56.25.

For Dataset 2, the results, as shown in Table \ref{tab:2a} are compared to eLDA; notably, eLDA achieves an average accuracy of 68.37\%, with 22 channels utilized in the classification process. On the other hand, the proposed framework attains an average accuracy of 66.68\% but demonstrates similar accuracy with a significant reduction in the number of selected channels to just one. Subject-specific results reveal variations in performance, suggesting that the proposed framework remains competitive with eLDA in classifying left and right despite its reduced channel selection. Later, a comparison of the proposed framework is also made with the best-performed model from another single-channel framework from the literature \cite{kanoga2018comparative}. Kanoga et al.,2018 select optimal channels among all the provided channels \cite{kanoga2018comparative}, whereas in our proposed framework we restrict it to be one among C3, Cz, and C4 across all datasets. From the comparison demonstrated in Table \ref{tab:2aa}, it is clear that the proposed framework outperforms PS-SVM \cite{kanoga2018comparative} with an average accuracy of 66.68 \% whereas the accuracy of PS-SVM\cite{kanoga2018comparative} being 63.5 \%.

Later the FRPC integrated framework was tested on Dataset 2 to examine all possible combinations of two class sets. Examining the results from Table \ref{tab:within}, it becomes evident that certain subjects exhibit optimal performance in specific tasks involving particular body parts. The average performance scores across all subjects demonstrate that the Right Hand - Tongue motor imagery set outperforms other body parts, with an average score of 72.8\%, followed closely by the Tongue-Feet set at 66.68\%. These findings suggest that, within the constraints of a single-channel system, motor tasks involving the Right Hand-Tongue set and Tongue-Feet set could be considered potential groups for optimized classification and targeted interventions. Understanding these performance variations is crucial for refining the capabilities of single-channel systems in classifying and interpreting motor tasks across different body parts.

 For Dataset 3, EEG data from 3 channels for 21 subjects was analyzed using the features extracted from Event-Related Synchronization/DeSynchronization, which are passed to the AdaBoost Classifier. These results are compared with the results achieved from the proposed framework. Table \ref{tab:21} shows variability across subjects, reflecting individual differences in neural responses. The average accuracy using ERD/S is calculated to be 61.21\%. The proposed framework outperforms with an accuracy of  67.6\%, emphasizing the importance of specific frequency bands from the best channel in optimizing the classification. These findings underscore the potential of the proposed framework in EEG-based neurofeedback applications, highlighting its ability to discern and classify neural activity patterns associated with different tasks.

\begin{table}
    \centering
    \caption{Comparision of Accuracies (in \%) for BCI IV-2b dataset}
    \begin{tabular}{ccccccc}
    \toprule
        \textbf{Subject ID} &\textbf{eLDA}  &   \textbf{Proposed FRPC} &  \textbf{Selected channel} \\
        & \cite{oikonomou2022multitask} &  \textbf{Framework} & \textbf{\& no. of bands} \\
            & [22 channels]& [1 channel] & \textbf{(ch,n)} \\
    \midrule
        B01 & 73.13  &  54.68&C3,6 \\
        B02& 61.79 &  51.56&C3,1\\
        B03 &  56.25&  58.12&C4,3\\
        B04 &  96.25&  80.39 & C4,2\\
        B05 &  91.25&  52.81& C3,6\\
        B06 &  82.19& 63.12  & Cz,6\\
        B07 &  73.75&  63.12& C3 ,6\\
        B08 &  91.56&  65.63 & C4,1\\
        B09 &  85.94 & 64.37 & C3,6\\
        \midrule
        Average & 79.12& 62.29  \\
 \bottomrule
    \end{tabular}
    \label{tab:2b}
    
\end{table}
\begin{table}
    \centering
    \caption{Comparision of Accuracies (in \%)for BCI IV-2a dataset}
    \begin{tabular}{ccccccc}
    \toprule
        \textbf{Subject ID} &\textbf{eLDA}  & \textbf{Proposed FRPC}  \\
        &  \cite{oikonomou2022multitask}&\textbf{Framework}  &  \\
            & [3 channels] & [1 channel]  \\
    \midrule
        A01 &  72.22  & 72.41 \\
        A02&  49.31  & 62.06 \\
        A03 &  86.11  &  72.41\\
        A04 &  58.33 &86.20 \\
        A05 &  73.61 &59.92 \\
        A06 &  56.25 & 60.68 \\
        A07 &  63.89 &  61.96\\
        A08 &  84.72 &61.7 \\
        A09 &  70.83  & 62.8 \\
        \midrule
        Average &68.3661 & 66.68 \\
    \bottomrule
    \end{tabular}
    \label{tab:2aa}
\end{table}
    
\begin{table}
    \centering
    \caption{Comparision of Accuracies(in \%) for BCI IV-2a dataset with other single channel framework}
    \begin{tabular}{ccccccc}
    \toprule
        \textbf{Subject ID}  &   \textbf{(PS-SVM)}& \textbf{Proposed FRPC} &  \textbf{Selected channel} \\
        &  \cite{kanoga2018comparative}&\textbf{Framework}  & \textbf{\& no. of  bands} \\
    \midrule
        A01 & 58.4 (CP4) & 72.41 & Cz,1\\
        A02&   56.1 FC2)  & 62.06 & C4,6\\
        A03 &     70.8(C3)  &  72.41& Cz,5\\
        A04 &    59.0 (POz) &86.20 & C4,3\\
        A05 &    55.6(CP1) &59.92  & C3,2\\
        A06 &   60.1 (FC4) & 60.68& Cz,1 \\
        A07 &    59.8(FC4) &  61.96& Cz,2\\
        A08 &    66.4 (CP4)  &61.7  & C3,5\\
        A09 &    85.0 (C4)  & 62.8 & Cz,1\\
        \midrule
        Average & 63.5 & 66.68& \\
    \bottomrule
    \end{tabular}
    \label{tab:2a}
\end{table}

\begin{table}
    \centering
    \caption{Comparision of Accuracies(in \%) for collected data}
    \begin{tabular}{ccccccc}
    \toprule
        \textbf{Subject} &  \textbf{ERD/S -AdaBoost}& \textbf{Proposed FRPC} &  \textbf{Selected channel} \\
        No.& & \textbf{Framework} & \textbf{\& no. of bands} \\
    \midrule
        01 &55  & 61.66 & C4,3\\
        02& 68.75 & 40 & C3,1\\
        03& 63.75& 82.5 & C4,5\\
        04& 58.75&  71.6& C3,6\\
        05& 56.25 &  68.33& C3,2\\
        06&  56.25&  69.16& C4,6\\
        07&  60&  66.66& C3,1\\
        08&  61.252&  61.66& Cz,6\\
        09&  67.52&  84.16& Cz,1\\
        10&  65& 80.83 & Cz,6\\
        11&  60& 64.16& C4,5\\
        12&  63.78& 71.66 & C4,4\\
        13&  65&  70.08& C4,2\\
        14&  56.25&  72.5& C3,6\\
        15&  61.25&  74.9& C3,2\\
        16&  71.25&  59.1& C4,6\\
        17&  63.75&  60& C3,1\\
        18&  55&  59& Cz,2\\
        19&  60&  72.5& C3,1\\
        20&  58.757&  75.83& Cz,1\\
        21&  58.75&  53.33& C3,2\\
    \midrule
    Average & 61.21&67.6 & \\
       
    \bottomrule
    \end{tabular}
    \label{tab:21}
\end{table}

\begin{table}
    \centering
    \caption{Comparision of Accuracies(in \%) for within-group BCI IV-2a dataset using Proposed FRPC Framework}
    \begin{tabular}{ccccccc}
    \toprule
        \textbf{Subject} &\textbf{Left-}  &  \textbf{Left-}&  \textbf{Left-}& \textbf{Right-} &  \textbf{Right-}&\textbf{Tongue-} \\
        \textbf{ID} &\textbf{Right}  &  \textbf{Feet}&  \textbf{Tongue}& \textbf{Feet} &  \textbf{Tongue}&\textbf{Feet} \\
    \midrule
        A01 &  72.4&68.9  &75.86  & 82.7 &93.1  & 82.7\\
        A02&  62.06&   55.17&41.37  & 51.72 &68.62 &65.51 \\
        A03 &  72.41&  65.51 & 72.41 &65.51 & 75.86 & 65.51\\
        A04 &  86.2&  68.96  &62.06  & 55.72&51.72 &58.62 \\
        A05 &  59.92& 64.46 & 64.39 & 60.15 & 67.87 &63.25 \\
        A06 & 60.68 & 64.92 &59.39  &58.1& 70.45 &58.1 \\
        A07 & 61.96 & 72.12 &73.71  &67.8  & 79.24 & 75.6\\
        A08 & 61.7 & 62.72 & 74.01 & 60.45 & 74.92 & 75.6 \\
        A09 & 62.8 &65.68  &85.9  &63.63  &79  &74 \\
        \midrule
        Average &  66.68&65.38  &67.67  & 62.86 &72.8&68.7   \\
        \bottomrule
    \end{tabular}
    \label{tab:within}
\end{table}

Combining all the 3 datasets used in this study, i.e. for a total of 39 subjects together, the most optimal channel mostly found is C3. Since the location of the C3 channel, according to 10-20, the EEG system is in the left hemisphere of the motor cortex, which is responsible for the dominant activity of the right part of the body, overall. This could be a reason for the strong occurrence of the channel, making C3 an optimal channel. This was followed by C4 and Cz being the least occurring channel. 

\section{Conclusion \& Future Work} 
\label{sec:discussion}
In conclusion, this study introduces an integrated framework for motor imagery classification with comprehensive analysis, employing a systematic approach to select the optimal channel and utilising Pearson correlation for filter bank selection, enabling the extraction of both spectral and temporal features. The efficacy of the proposed framework is rigorously tested on two publicly available datasets and one collected dataset, demonstrating its versatility across diverse datasets. Notably, the framework's performance is benchmarked against eLDA for Dataset 1\& 2 and single channel PSD-SVM-based framework for Dataset 2, showcasing its competitive edge in 2-class motor imagery classification tasks. Beyond the implementation of the framework, this study delves into the identification of C3 as the most optimal channel across all subjects, which has been repeated more frequently, enhancing our understanding of individual-specific neural responses.  Additionally, the exploration of classes where the single-channel framework exhibits an efficiency of the proposed framework being efficient for Tongue-Feet based Motor Imagery classification. Future work involves integrating the pipeline into 1D-CNN and also exploring other classifiers like SVM, LDA, etc.

\bibliographystyle{IEEEtran} 
\bibliography{bib} 
\vspace{12pt}
\color{red}
\end{document}